\documentclass[twocolumn,showpacs,amsmath,amssymb]{revtex4}
\usepackage{graphicx}

\begin{document}



\title{\bf A repair strategy to the attacked random and scale-free networks}

\author{L.P. Chi$^\dag$, C.B. Yang and X. Cai}
\affiliation{Institute of Particle Physics, HuaZhong Normal
University, Wuhan 430079, China \\
$^\dag$Email: chilp@iopp.ccnu.edu.cn\\
}

\date{\today}

\begin{abstract}

With a simple attack and repair evolution model, we investigate
the stability and structural changes of the Erd\"{o}s-R\'{e}nyi
random graphs (RG) and Barab\'{a}si-Albert scale-free (SF)
networks. We introduce a new quantity, \textit{invulnerability}
$I(s)$, to describe the stability of the system. We find that both
RG and SF networks can evolve to a stationary state. The
stationary value $I_{c}$ has a power-law dependence on the repair
probability $p_{re}$. We also analyze the effects of the repair
strategy to the attack tolerance of the networks. We observe that
there is a threshold, $(k_{max})_c$, for the maximum degree. The
maximum degree $k_{max}$ at time $s$ will be no smaller than
$(k_{max})_c$. We give further information on the evolution of the
networks by comparing the changes of the topological parameters,
such as degree distribution $P(k)$, average degree $\langle k
\rangle$, shortest path length $L$, clustering coefficient $C$,
assortativity $r$, under the initial and stationary states.

\end{abstract}


\maketitle

\section{Introduction}

Most of the systems in the real world can be represented as
complex networks \cite{s1, s2, s3, s4, s5, s6, s7,s8}, in which
the nodes are the elementary components of the system and the
edges connect pair of nodes that mutually interact exchanging
information. The existing empirical and theoretical results
indicate that complex networks can be divided into two major
classes based on their degree distribution $P(k)$ \cite{s10}. The
degree of a node is the number of links adjacent to the node and
the degree distribution gives the probability that a node in the
network is connected to $k$ other nodes. The first class of
networks is characterized by Poisson degree distribution, such as
Erd\H{o}s and R\'{e}nyi random graph \cite{s11}. It is constructed
starting from an initial condition of $N$ nodes and then adding
links with connection probability $p$. The second class of
networks is characterized by power-law degree distribution $P(k)
\sim k^{-\gamma}$, such as Barab\'{a}si and Albert scale-free
network \cite{s12}. It starts from an initial condition of $m_0$
nodes and then, for each time step, evolves adding a new node with
$m$ edges (growth) that is connected more likely to nodes with
higher degree (preferential attachment).

Recently enormous interest has been devoted to study the
robustness of complex networks under random failures or
intentional attacks \cite{s13,s14,s15,s16,s17,s18}. By random
failure we mean the removal of randomly selected network nodes and
their edges. Instead, we call intentional attack the targeted
removal of the most important nodes and their edges. The
importance of a node can be determined by the degree, or the
betweenness (the number of shortest paths that pass through the
node), \textit{etc}.

In this paper, the node with highest degree suffers attack. Since
this node has highest connections to other nodes, it is most
likely attacked and is the most vulnerable site in the network.
The aim of this paper is to investigate and compare the stability
of Erd\H{o}s and R\'{e}nyi random graph and Barab\'{a}si and
Albert scale-free network under the attack and repair strategy. At
the same time, we will compare the effects of the attack and
repair strategy to these two different topological structures. The
paper is organized as follows. In Section II, we describe the
evolution model of the attack and repair strategy. Simulation
results are in the next section. In the conclusion, we formulate
the main results of the paper.

\section{The attack and repair strategy}

Our attack and repair strategy is simulated as follows. We start
by constructing a network according to the algorithms of Erd\H{o}s
and R\'{e}nyi random graph (RG) or Barab\'{a}si and Albert
scale-free network (SF). In RG, the network size is $N$ and the
connection probability is $p$. In SF networks, the initial $m_0$
and $m$ are both 2. The dynamics of our model is defined in terms
of the following two operations.

\begin{enumerate}

\item[(i)] \textit{Attack}: Find a node with the
maximum degree $k_{max}$ and remove all
its links. (If several nodes happen to have the same highest
degree of connection, we randomly choose one of them.)

\item[(ii)] \textit{Repair}: Reconnect this node
with the other nodes in the network with probability $p_{re}$. We
assume that all information on the former links of the damaged
nodes has been lost, the damaged nodes have be connected randomly
to other nodes in the network again.

\end{enumerate}

Then, the evolution comes into the next Monte Carlo time step. An
issue should be mentioned here. Unlike previous attack model
\cite{s12} the nodes damaged will not be removed from the network,
instead they will be reconnected in certain way. In this way, we
keep the size of the system constant.

\section{simulation results}

\begin{figure}
\begin{center}
\includegraphics[width=7cm]{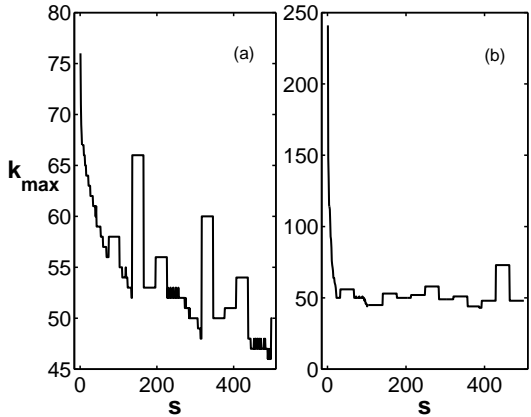}
\caption{Plots of $k_{max}$ versus time step $s$ with size
$N=10000$ and $p_{re}=0.005$ for (a) random graphs; (b) scale-free
networks.}
\end{center}
\label{fig1}
\end{figure}

We have performed extensive numerical simulations to study the
evolution of RG and SF networks under the attack and repair
strategy. In Fig. 1, we give a snapshot of the maximum degree
$k_{max}$ versus time step $s$ with $N=10000$ nodes and repair
probability $p_{re}=0.005$. We find that $k_{max}$ of RG model
decreases slowly with big fluctuations during the evolution. For
SF model, $k_{max}$ decreases steeply at the beginning of the
evolution and then slowly later. This behavior is rooted in their
different topological structures. The RG network is a homogenous
network and all nodes have approximately the same number of links.
Thus the removal of each node causes the same amount of damage.
The SF networks is an extremely inhomogeneous network, in which
the minority of nodes have most of the links. The removal of these
'important' nodes will have dramatic impact on the changes of
$k_{max}$.

Generally, the systems with larger $k_{max}$ have higher
probability to be attacked. In order to evaluate the probability
that the system is vulnerable, we introduce the concept,
\textit{invulnerability}, $I(s)$. Considering an evolution of
network with maximum degrees $k_{max}(1)$, $k_{max}(2)$,...,
$k_{max}(s)$, invulnerability $I(s)$ at time $s$ is defined as

\begin{equation}
I(s)=1/Min\{k_{\rm max}(i)\} \ \ \textbf{for\ \ \ } i\leq s\ ,
\label{invul}
\end{equation}

\noindent According to Eq.~(\ref{invul}), $I(s)$ is a
non-decreasing parameter, and helps filter out some fluctuations
in $k_{max}$.

\begin{figure}
\begin{center}
\includegraphics[width=7cm]{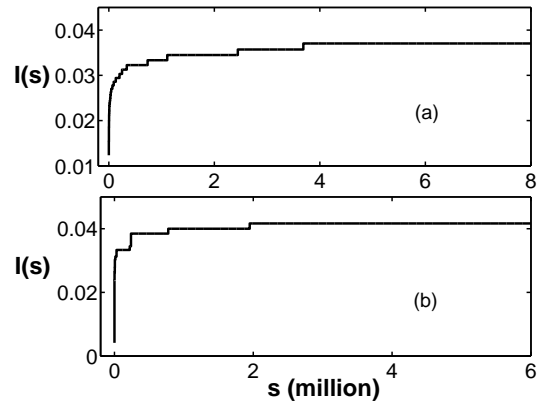}
\caption{Plots of $I(s)$ versus step $s$ with $N=10000$ and
$p_{re}=0.005$ for (a) random graphs; (b) scale-free networks.}
\end{center}
\label{fig2}
\end{figure}

In Fig. 2 we plot the evolution of $I(s)$ versus $s$ for RG and SF
networks under $N=10000$ and $p_{re}=0.005$. We observe that in
both networks $I(s)$ increases very quickly at small $s$ but
slowly at large $s$ and finally reaches a stationary value $I_{c}$
after taking a long transient period. The stationary value $I_c$
of the RG network is 0.037, while $I_c$ of the SF network is
0.042.

Additionally, we can see from Eq.~(\ref{invul}) that the maximum
degree $k_{max}$ of the network in the stationary state is

\begin{equation}
(k_{max})_{c}=\frac{1}{I_c}.
\label{kc}
\end{equation}

\noindent Since stationary value $I_c$ is the maximum of all the
$I(s)$, $k_{max}$ at time $s$ will be no smaller than
$(k_{max})_c$, that is, $k_{max}(s)\geq (k_{max})_c$. For
instance, under $N=10000$ and $p_{re}=0.005$, the repair strategy
ensures the maximum degree $k_{max}$ of the RG network is no
smaller than $1/I_c=1/0.037=27$ and the $k_{max}$ of the SF
network is no smaller than $1/I_c=1/0.042=24$. The introduction of
the repair strategy ensures that $k_{max}$ of the attacked RG and
SF networks are greater than 27 and 24 during the evolution,
respectively. If without the repair strategy, all the links will
be removed after about 5000 time steps for attacked RG model, 2000
time steps for attacked SF model.

\begin{figure}
\begin{center}
\includegraphics[width=6cm]{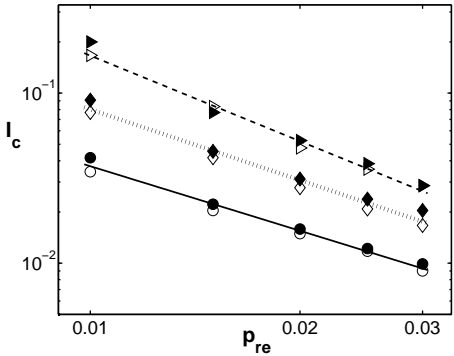}
\caption{Plot of $I_{c}$ versus repair probability $p_{re}$ with
network size $N=2000$ (triangles); 3000 (squares); and 5000
(circles) for RG and SF networks.}
\end{center}
\label{fig3}
\end{figure}

In order to understand the effects of the repair strategy to the
stationary state of the system, we plot the stationary value $I_c$
as a function of the repair probability $p_{re}$ in log-log scale
for RG networks (open symbols) and SF networks (filled symbols)
with different network sizes. Triangles, diamonds and circles
represent $N=2000$, $3000$ and $5000$, respectively. We find that
under fixed network size $N$, the higher the repair probability
$p_{re}$ is, the smaller the stationary value $I_c$ is. On the
other hand, under the fixed $p_{re}$, the larger $N$ is, the
smaller $I_c$ is. In fact, $I_c$ has an inverse ratio to the
product of $N$ and $p_{re}$.

Figure 2 also tells that the stationary value $I_c$ has a
power-law dependence on the repair probability $p_{re}$,

\begin{equation}
I_c \sim p_{re}^{-\tau}.
\end{equation}

\noindent Under the same network size $N$ and repair probability
$p_{re}$, the stationary values $I_c$ of RG and SF networks are
very similar, with $I_c$ of SF networks a little larger, thus more
invulnerable.

By investigating the changes of the topological parameters of RG
and SF networks at the initial and stationary states, we are
trying to understand the effects of the repair strategy to the
topological structures of these two different structural networks.

\textit{Degree distribution}. Figure 4(a) and 4(b) display the
degree distributions $P(k)$ of RG and SF networks in the initial
and stationary states under $N=5000$ and $p_{re}=0.02$. Figure
4(a) and 4(b) show that the shape of the degree distributions is
almost the same for RG and SF networks in initial and stationary
states. For RG networks, the degree distributions in the initial
and stationary state are both Gaussian distributions, with the
peak changes from 100 in the initial state to 38 in the stationary
one. At the same time, the distribution of SF networks in the
initial and stationary states are both power-law distribution,
having the same exponent about 2.6. Figure 4 suggests that the
general structures of the systems do not change even after a long
time evolution to the stationary states.

\begin{figure}
\begin{center}
\includegraphics[width=7cm]{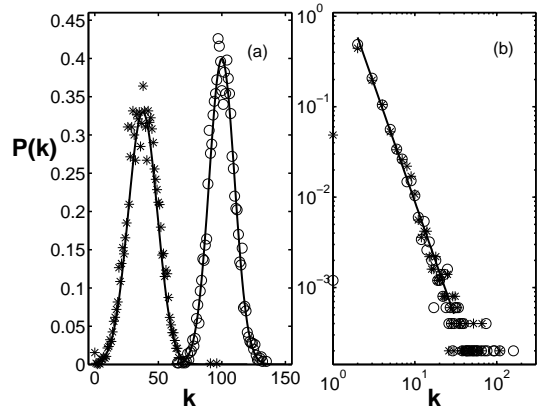}
\caption{A comparison of degree distributions under $N=5000$ and
$p_{re}=0.02$ in initial (circles) and stationary (pluses) states
for (a) random graphs; and (b) scale-free networks. Solid lines
are the fit curves. }
\end{center}
\label{fig4}
\end{figure}

\textit{Average degree}. We find that the average degree of RG
network in the stationary state $\langle k \rangle$ decreases
dramatically, while slightly for SF networks. This result is
caused by the homogeneity of RG networks. Since in RG networks all
nodes have approximately the same number of links, majority of
nodes in the network will be impacted by the attacks and, hence,
the rapid decrease of $\langle k \rangle$. The power-law topology
of SF networks displays that $\langle k \rangle$ is mainly
determined by the nodes with small degrees, which are less
influenced by attacks.

\textit{Shortest path length}. The shortest path length $L$ is the
average length of the shortest paths between any two nodes in the
network. $L$ in the stationary states for RG and SF networks both
increases. That is, the efficiency for the nodes in the network to
communicate decreases under the attacks.

\textit{Clustering coefficient}. The clustering coefficient of
node $i$ is defined as the existing numbers $n_{i}$ among the
links of node $i$ over all the possible links, that is,
$C_{i}=n_{i}/[k_{i}(k_{i}-1)/2]$. The clustering coefficient of
the whole network is the average over all the $C_i$. The increase
of clustering coefficient for RG networks in the stationary states
means the nodes have the tendency to be highly clustered. In fact,
the small shortest path length and large clustering coefficient
suggest the RG network in the stationary states is more or less a
small-world network. The clustering coefficient of SF networks
decreases a little in the stationary states. It seems that few
nodes with largest connections are attacked and have a big affect
on the network, which makes the connections in the network becomes
sparse, thus the clustering coefficient becomes decrease.

\textit{Assortativity}. The assortativity $r$ in the range $-1
\leq r \leq 1$ is another interesting feature of complex networks.
A network is said to show assortative, $r>0$, if the high-degree
nodes in the network preferentially connect to other high-degree
nodes. By contrast, a network is disassortative, $r<0$, if the
high-degree nodes tend to connect to other low-degree nodes. The
increase of $r$ for RG networks in the stationary states are
consistent with our conjecture that the networks becomes a little
highly clustered. For SF networks, they are inhomogeneous in both
initial and stationary states according to the analysis of the
degree distribution above, so $r$ are both negative, that is
disassortative.

\begin{table}[t]
\caption{\em Comparison of topological changes in initial and
stationary states between RG and SF networks under $N=5000$ and
$p_{re}=0.02$.}
\begin{center}
\begin{tabular}{c|c|c|p{0.15cm}|c|c}
\hline\hline
&\multicolumn{2}{c|}{\bf{RG}}  & & \multicolumn{2}{|c}{\bf{SF}}           \\
\cline{2-3}\cline{5-6}
                    & initial & stationary  & & initial & stationary\\
\hline
$I_{c}$             & ---     &  0.0149  & & --- & 0.0152   \\
$\langle k \rangle$ & 100.06   & 38.09  & &  3.99     & 3.92 \\
$ L $               & 2.11    & 3.89  & &   4.83    & 5.28\\
$ C $               & 0.02   & 0.47  & &  0.0068  & 0.0028\\
$ r $               & 0.23 & 0.52  & &   -0.062  & -0.074\\
\hline

\end{tabular}
\end{center}
\label{comparison}
\end{table}

\section{Conclusions}

Summarily, in this paper we study the effects of the repair
strategy to the attacked ER random graph model and BA scale-free
model. We find the introduction of the repair strategy has
significant effects on the attack tolerance of the networks. We
also introduce a new quality, \textit{invulnerability} $I(s)$, to
describe the stability of the networks. After a long time
evolution, the invulnerability of ER random graph model and BA
scale-free model can both evolve to a stationary value $I_c$.
$I_c$ has a power-law dependence on the repair probability
$p_{re}$. In addition, the maximum degree in the stationary state
$(k_{max})_c$ is always smaller than the maximum degree at time
$s$, $k_{max}(s)$.

We give further information on the evolution of the networks by
comparing the changes of the topological parameters, such as
degree distribution $P(k)$, average degree $\langle k \rangle$,
shortest path length $L$, clustering coefficient $C$ and
assortativity $r$, between the initial and stationary states. We
conclude that the topological structure of SF networks does not
change during the evolution, while RG networks become small-world
networks during the evolution, according to the small shortest
path length and large clustering coefficient.

\section*{ACKNOWLEDGEMENTS}

This work was supported by National Natural Science Foundation of
China under Grant Nos. 70401020, 70571027 and 10647125.

\end{document}